\documentclass{aa}
\usepackage{epsfig}
\usepackage{txfonts}

\usepackage{epsfig}

\newcommand{\zsun}{\mbox{$Z_{\odot}$}}

\newcommand{\kms}{km s$^{-1}$}

\newcommand{\degree}{$^{\rm o}$}

\newcommand{\bb}{\bibitem[]{bla}}

\def\lesssim{\mathrel{\hbox{\rlap{\hbox{\lower4pt\hbox{$\sim$}}}\hbox{$<$}}}}
\def\gtrsim{\mathrel{\hbox{\rlap{\hbox{\lower4pt\hbox{$\sim$}}}\hbox{$>$}}}}

\def\arcsec{\hbox{$^{\prime\prime}$}}

\begin{document}


\title{Constraining GRB progenitor models by probing Wolf-Rayet wind geometries in 
the Large Magellanic Cloud}

\author{Jorick S. Vink\inst{1,2}}
\offprints{Jorick S. Vink, jsv@arm.ac.uk}

\institute{Keele University, Lennard Jones Laboratories, ST5 5BG, United Kingdom \and
          Armagh Observatory, College Hill, Armagh BT61 9DG, Northern Ireland, United Kingdom}

\titlerunning{WR wind geometries in the LMC}
\authorrunning{Jorick S. Vink}

\abstract{The favoured progenitors of long-duration gamma-ray bursts (GRBs) are rapidly rotating 
Wolf-Rayet (WR) stars. However, most Galactic WR stars are slow rotators, as stellar winds are thought to remove 
angular momentum. This poses a serious challenge to the collapsar model. Recent observations indicate that GRBs occur predominately in low 
metallicity ($Z$) environments, which may resolve the problem: lower $Z$ leads to less mass loss, which may inhibit  
angular momentum removal, allowing WR stars to remain rotating rapidly until collapse.}
{We wish to determine whether low $Z$ WR stars rotate on average more rapidly than Galactic WR stars.}
{We perform a Very Large Telescope (VLT) linear spectropolarimetry survey of WR stars in 
the low $Z$ environment of the Large Magellanic Cloud (LMC) and compare our results with the Galactic 
sample of Harries et al. (1998).}
{We find that only 2 out of 13 (i.e. 15\%) of LMC WR stars show line polarization effects, compared to a similarly low fraction 
of $\sim$ 15-20\% for Galactic WR stars.}
{The low incidence of line polarization effects in LMC WR stars suggests that the threshold metallicity where significant 
differences in WR rotational properties occur is below that of the LMC ($Z$ $\sim$ 0.5 $\zsun$), possibly constraining GRB progenitor channels 
to this upper metallicity.}

\keywords{Stars: Wolf-Rayet -- Stars: early-type -- Stars: mass-loss
          -- Stars: winds, outflows -- Stars: evolution}

\maketitle


\section{Introduction}
\label{s_intro}

A picture is emerging in which long-duration gamma-ray bursts (GRBs) are associated with 
the endpoints of the life of a massive star 
(Galama et al. 1998; Hjorth et al. 2003; Stanek et al. 2003). This has 
given huge impetus to the collapsar model for gamma-ray bursts of Woosley (1993).
The model relies on both the progenitor being rapidly rotating, allowing it  
to subsequently collapse to a black hole, as well as the progenitor being a relatively 
compact star that has lost its hydrogen (H) envelope, allowing the jet to emerge. 
The progenitor is therefore generally considered to be a H-poor Wolf-Rayet (WR) star 
(e.g. Mirabal et al. 2003).

The challenge of these two requirements is that they seem mutually exclusive. 
Massive stars posses strong stellar winds,
which may remove the angular momentum (Langer 1998; Maeder \& Meynet 2000), 
leaving a slowly rotating WR star before collapse. 

However, the winds of massive stars are predicted to scale with the 
metallicity (Abbott et al. 1982, Kudritzki et al. 1987, Leitherer et al. 1992, 
Vink et al. 2001), which would imply that stars in lower metallicity environments 
could rotate faster.
A very relevant development in the predictions of the 
metallicity dependence of stellar winds has been that WR mass loss should scale 
with the Fe content of the host galaxy (Vink \& de Koter 2005; Gr\"afener \& Hamann, in prep.), 
and not with self-enriched species -- as previously often assumed (e.g. Maeder \& Meynet 2003). 
The steep scaling of WR mass loss with $\dot{M}$ proportional to 
$Z^{0.6-0.8}$, as predicted by Vink \& de Koter (2005) appears to be confirmed 
by observations (Crowther 2006).
The metallicity dependence of WR stars is also backed up with stellar evolution models that 
can account for the drop in WC/WN ratio in galaxies at low $Z$ (Eldridge \& Vink 2006; see also 
Van Bever \& Vanbeveren 2003). 

Another intriguing aspect is that the inferred steepness of the WR mass loss $Z$ dependence may explain 
the inhibition of angular momentum removal from WR stars at low $Z$, presenting a boost to 
the collapsar model for GRBs at low metal content (Yoon \& Langer 2005; Woosley \& Heger 2006).
Therefore, the earlier mentioned paradox for the explanation of long-duration GRBs from 
rapidly rotating WR stars may be resolved if they would occur at low $Z$. Intriguingly, recent 
observations indeed indicate that 
GRBs are favoured in regions of low metallicity (e.g. 
Fynbo et al. 2003; Prochaska et al. 2004; Vreeswijk et al. 2004).

In this paper, we test the physical criteria of GRB progenitors observationally. 
If the WR mass-loss metallicity dependence and the subsequent inhibition 
of angular momentum removal are indeed the {\it key} to account for the predominant occurrence of GRBs at 
low $Z$, WR stars in the Magellanic Clouds should be rotating more rapidly than those in the Galaxy. 
We therefore wish to infer the fraction of rapidly rotating WR stars in the low-$Z$ environment of the LMC 
and compare it to that in the Galaxy. Unfortunately, direct v sin$i$ measurements are not feasible for WR stars 
but rapid rotation is believed to induce wind asymmetries (e.g. Maeder \& Meynet 2007) 
which can be probed with linear spectropolarimetry.

Despite this tool being ``photon-hungry'', linear spectropolarimetry is a 
powerful technique to measure asymmetries, as one of 
the basic principles is the following: continuum light is polarized by Thomson scattering off 
free electrons. For the case of a spatially unresolved spherical wind, polarized photons 
from all directions cancel, leaving no net polarization signal.
In case the wind departs from spherical symmetry, a net continuum 
polarization should be detected.     
{\it Line} photons are formed by recombinations over a larger volume and scatter and 
polarize less than the continuum photons. As a result, the line 
becomes depolarized with respect to the continuum -- if the wind is asymmetric. 

The method has been extensively used for rapidly rotating Be stars. 
Poeckert \& Marlborough (1978) found $\sim$ 60\% of Be stars to be intrinsically polarized, which is, due to random orientation of 
rotational axes, consistent with {\it all} Be stars having asymmetric outflows induced by rapid rotation. 
In a similar vein, Harries et al. (1998) performed a linear spectropolarimetry survey 
on 16 Galactic WR stars and found line effects in only 15--20 \% of them.
Their results suggested that only a small fraction of Galactic WR stars have significant asymmetric outflows due to rapid rotation.

Here, we present a {\it Very Large Telescope (VLT)} linear spectropolarimetry survey on a 
complete sample of the brightest LMC WR stars (with $V \la 12.3$) and we compare our results 
to the Galactic sample. If WR stars at low $Z$ indeed remain rapidly rotating, we may measure a higher 
fraction of line depolarizations.


\section{Observations}
\label{s_obs}

Our targets were selected from the fourth catalogue of Population I WR stars in 
the LMC by Breysacher, Azzopardi \& Tester (1999; hereafter BAT) on the basis of their 
relative brightness ($V \la 12.3$). 
Our LMC sample-size (of 13) is comparable to that from the Galactic 
WR spectropolarimetry study by Harries et al. (1998).
We note that targets were not chosen on the basis of any known circumstellar geometries.
As we currently know little about the polarimetric evolution 
of WN (nitrogen rich) into WC (carbon rich) into WO (oxygen rich) stars, nor do we know which of these subgroup WR stars 
may give rise to GRBs, we first wish to establish the overall difference 
in rotationally induced wind asymmetry between the LMC and the Galaxy. 
The list of objects is given in Table~\ref{t_cont}, alongside their {\it V} magnitudes and 
spectral types.

The linear spectropolarimetry data were obtained during the 
nights of 2006 October 29 and 30 using the FORS1 spectrograph on the VLT UT2 (Kuyen) 
in PMOS mode. The exposure times are given in column (4) of Table~1. 
The observations were obtained in a very similar manner to that of the Herbig Ae/Be 
study by Vink et al. (2002) to which we refer for details.

We used the GRIS600R+14 grism (and the GG435+31 order filter) with a slit width of 0.51$\arcsec$, which 
yielded a resolution of approximately 3\AA, i.e. 150 km/s around 6500\AA, i.e. a very similar setting 
to the Galactic WHT study by Harries et al. (1998). Due to relatively poor seeing 
we increased the slit width to 0.8$\arcsec$ during parts of the run.
Depolarizations can be measured across the 
He {\sc ii} line at 6560\AA\ and several other emission lines, such as the 
He {\sc ii} line at 5976\AA\ and the C{\sc iv} line at 5801\AA\ for WC stars. 
As WR stars have fast winds (of order $\sim$ 1500 -- 5000 \kms), we are able 
to resolve these lines. 

\begin{figure*}
\mbox{
\epsfxsize=0.37\textwidth\epsfbox{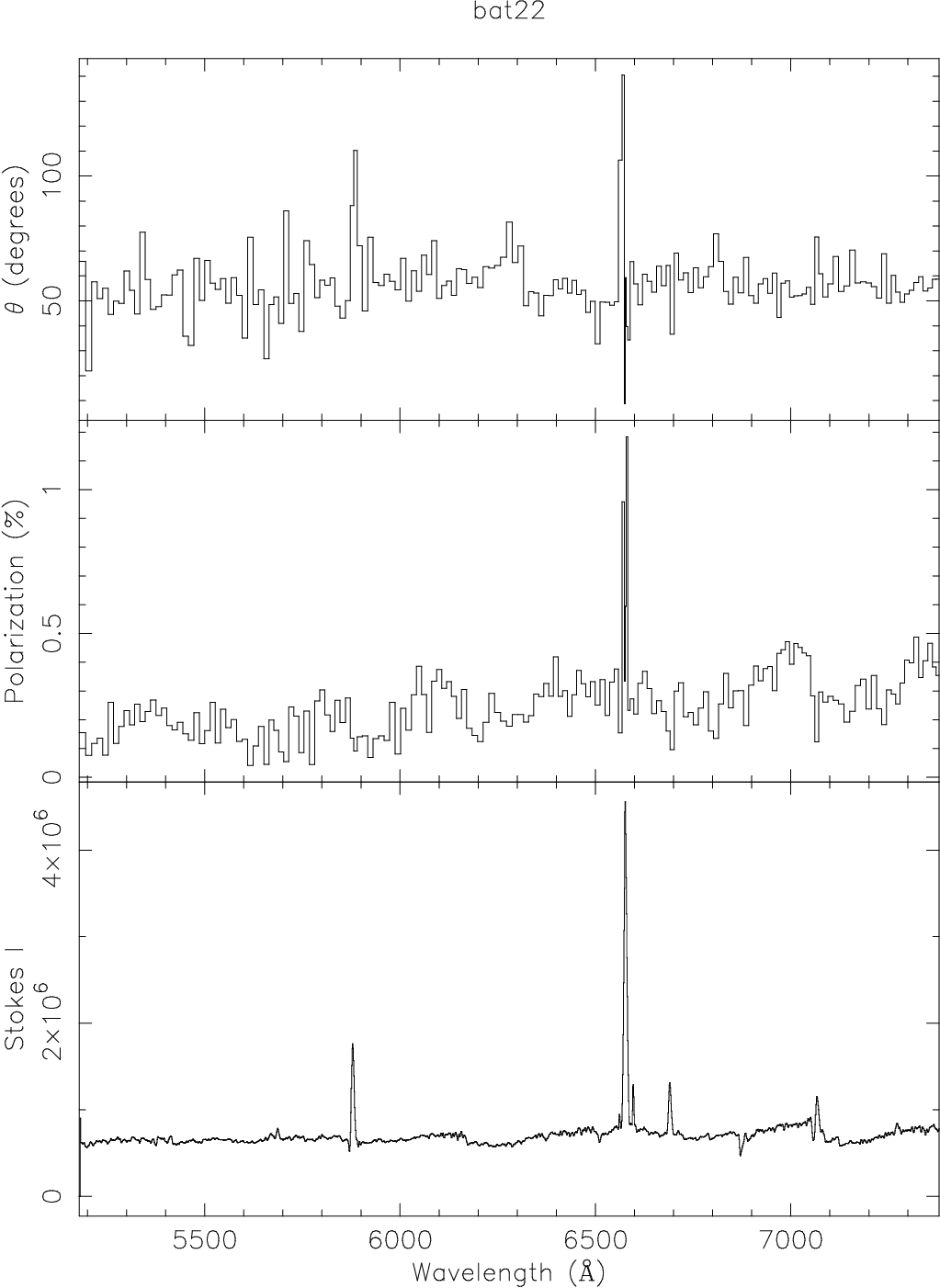}
\epsfxsize=0.37\textwidth\epsfbox{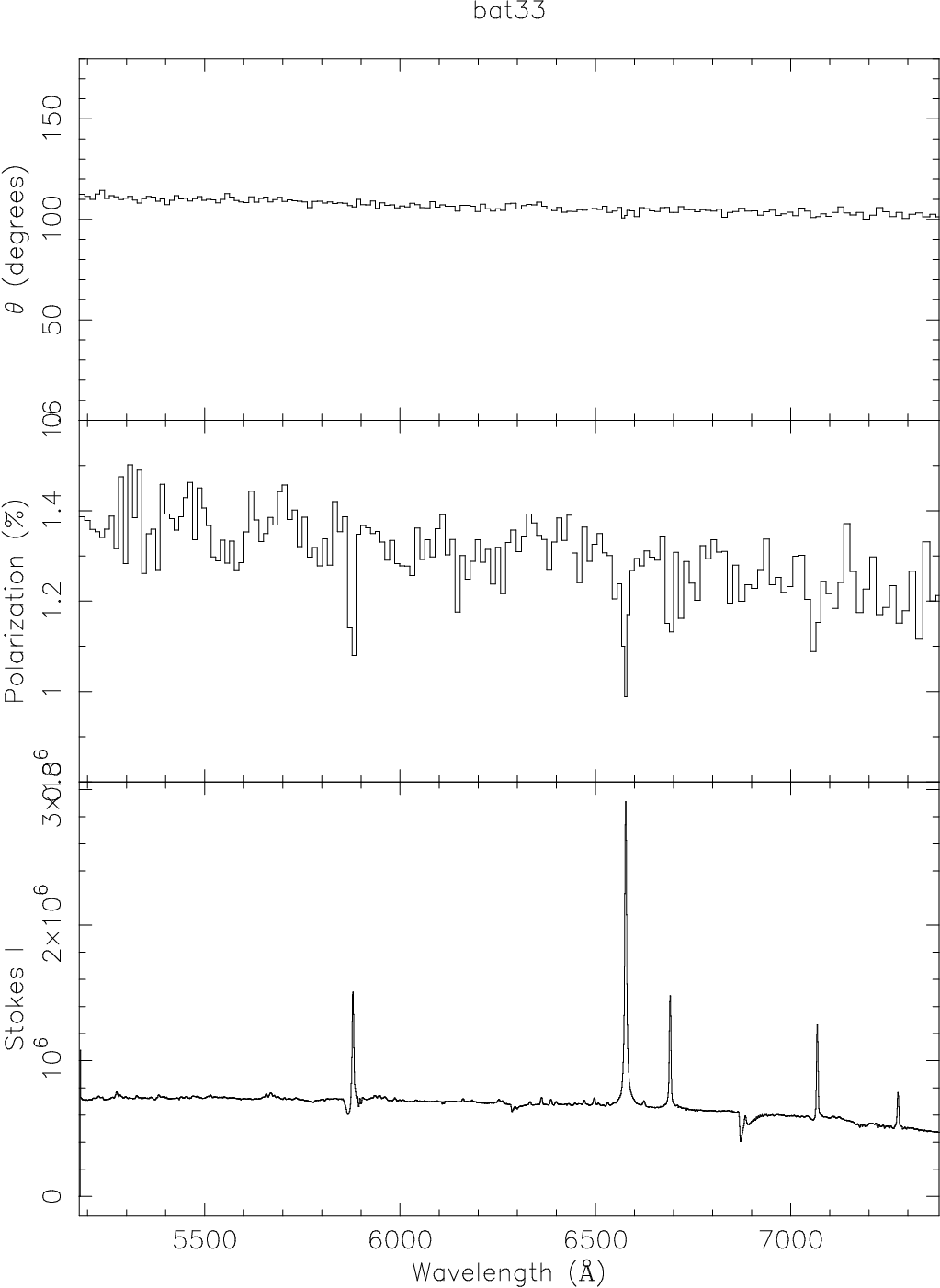}}
\caption{Polarization triplots for the WR targets that 
show depolarization. 
The Stokes I spectra are shown in the lower panel, 
the \%Pol are shown in the middle panel, and 
the position angles are plotted in the upper panel of 
the triplots.
}
\label{f_line}
\end{figure*}

To analyze the linearly polarized component in the spectra, 
FORS1 was equipped with the appropriate polarization optics.
Polarization and un-polarized standard stars were also observed.
The data reduction was carried out using {\sc iraf} for the usual 
bias and sky-subtraction, cosmic ray removal, spectrum extraction and 
wavelength calibrations of the (extra)ordinary rays, and  
{\sc starlink} software to determine $Q$ and $U$, as well as   
the percentage linear polarization $P$ and its position angle $\Theta$.
The achieved accuracy of the polarization data is in principle determined by 
photon-statistics only, which can be rather good ($<$ 0.01\%; see Table~1). However due to 
systematic effects, the absolute accuracy is lower. We do not correct 
for interstellar polarization (ISP) as these are equal for line and continuum 
wavelengths, but we note that interstellar polarization may act in a way 
that depolarization of the source turns into an enhancement
of $P$ (and an accompanying change in $\Theta$) in the data (see e.g. BAT 22 below).

\section{Results}
\label{s_res}

The continuum polarization percentage and its position angle are measured over 
a region between 5900\AA\ and 6500\AA\ and given in columns (5) and (6) of Table 1.
Figures~1 and~2 show the polarization data of all targets 
in the form of triplots. The lower panels show the 
Stokes I ``intensity'' data, the linear polarization percentages are plotted in the middle panels and 
the position angles are in the upper triplot panels. The 
polarization data are binned to a constant error of 0.05\%. 

\begin{table}
\begin{minipage}{\linewidth}
\renewcommand{\thefootnote}{\thempfootnote}
\caption{WR targets. The {\it V}-band magnitudes are taken from 
from Breysacher et al. (1999, BAT) and listed in column (2). 
The Spectral types (column 3) are taken from BAT (see references therein).
The integration times (column 4) denote the total {\it on source} exposures. 
The observed continuum position angle ($\Theta$) and 
its error are measured between 5900 \AA\ and 6500 \AA\ and given in columns (5) and (6).}
\label{t_cont}
\begin{tabular}{lclrcc}
\hline
BAT  & {\it V} & Spec            & Exp(s) & $P_{\rm cont}$ (\%) & $\Theta_{\rm cont}$ (\degree)\\
\hline
22 & 12.09 &  WN9h    & 1600        &  0.235 $\pm$ 0.007      &  147.8 $\pm$ 0.9              \\       
33 & 11.54 &  WN9       & 1400        &  1.315 $\pm$ 0.007      &  106.2 $\pm$ 0.2              \\
\hline
27 & 11.31 &  WN5?b   & 1400        &  0.254 $\pm$ 0.006      &   36.3 $\pm$ 0.7               \\
28 & 12.23 &  WC6     & 2400        &  1.038 $\pm$ 0.007      &   47.5 $\pm$ 0.2               \\
38 & 11.50 &  WC4    & 1600        &  0.556 $\pm$ 0.007      &   21.1 $\pm$ 0.3               \\
39 & 12.51 &  WC4    & 1000        &  0.445 $\pm$ 0.013      &   23.0 $\pm$ 0.9               \\
42 &  9.91 &  WN5?b    &  360        &  0.601 $\pm$ 0.007      &   24.5 $\pm$ 0.4                \\
  55 & 11.99 &  WN11h          & 1200        &  0.226 $\pm$ 0.009      &   30.7 $\pm$ 1.1           \\
  85 & 11.75 &  WC4        & 1600        &  1.716 $\pm$ 0.007      &  104.1 $\pm$ 0.1             \\
  92 & 11.51 &  WN6      & 1200        &  1.043 $\pm$ 0.006      &   81.6 $\pm$ 0.2             \\
 107 & 12.12 &  WNL         & 2400        &  1.647 $\pm$ 0.006      &   70.1 $\pm$ 0.1             \\
 118 & 11.15 &  WN6h           & 1200        &  0.166 $\pm$ 0.006      &   20.6 $\pm$ 1.1               \\
 119 & 12.16 &  WN6(h)         & 2400        &  2.231 $\pm$ 0.006      &   81.2 $\pm$ 0.1               \\
\hline
\end{tabular}
\\
\noindent
\end{minipage}
\end{table}

\begin{figure*}
\mbox{
\epsfxsize=0.33\textwidth\epsfbox{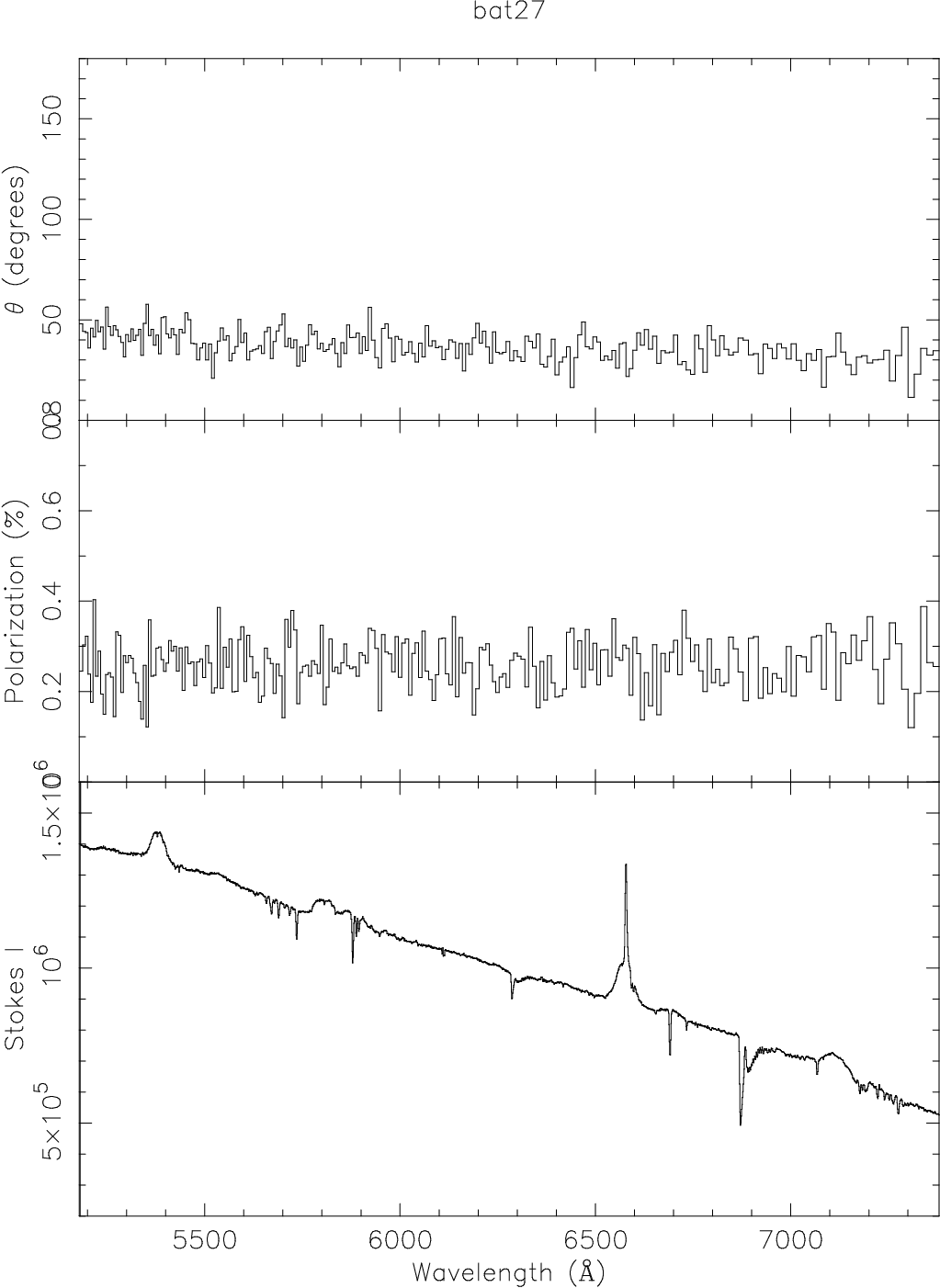}
\epsfxsize=0.33\textwidth\epsfbox{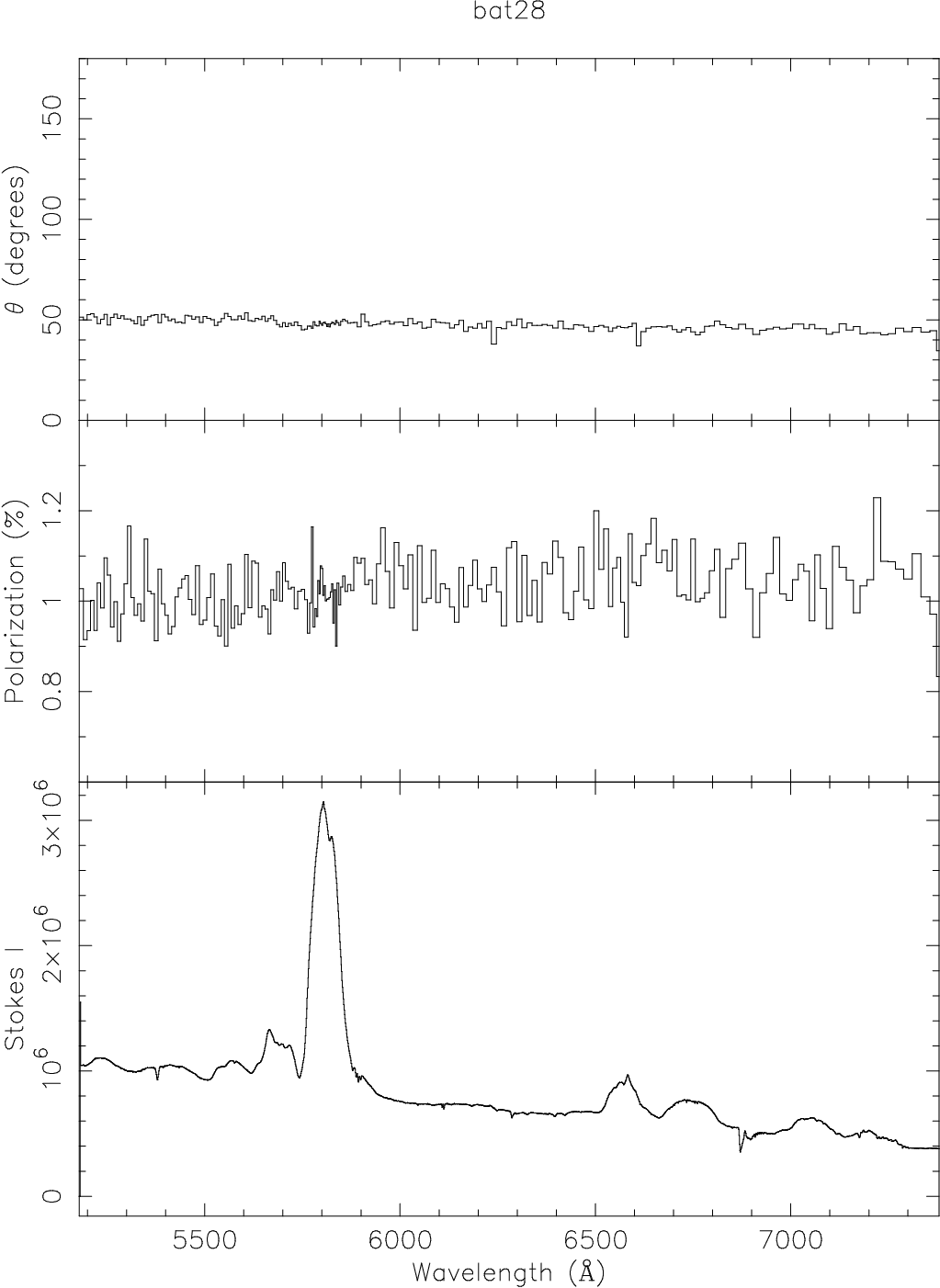}
\epsfxsize=0.33\textwidth\epsfbox{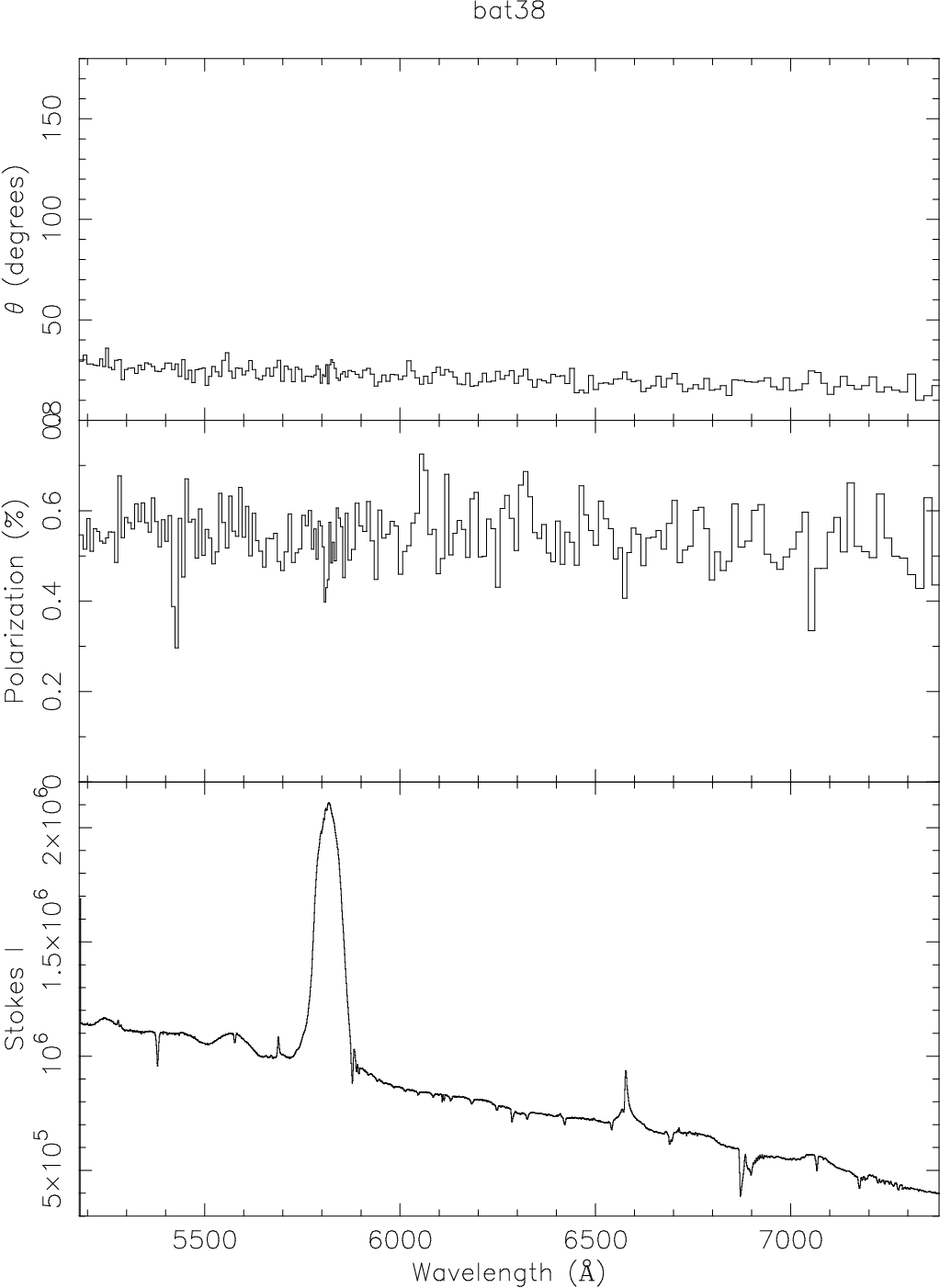}
}
\mbox{
\epsfxsize=0.33\textwidth\epsfbox{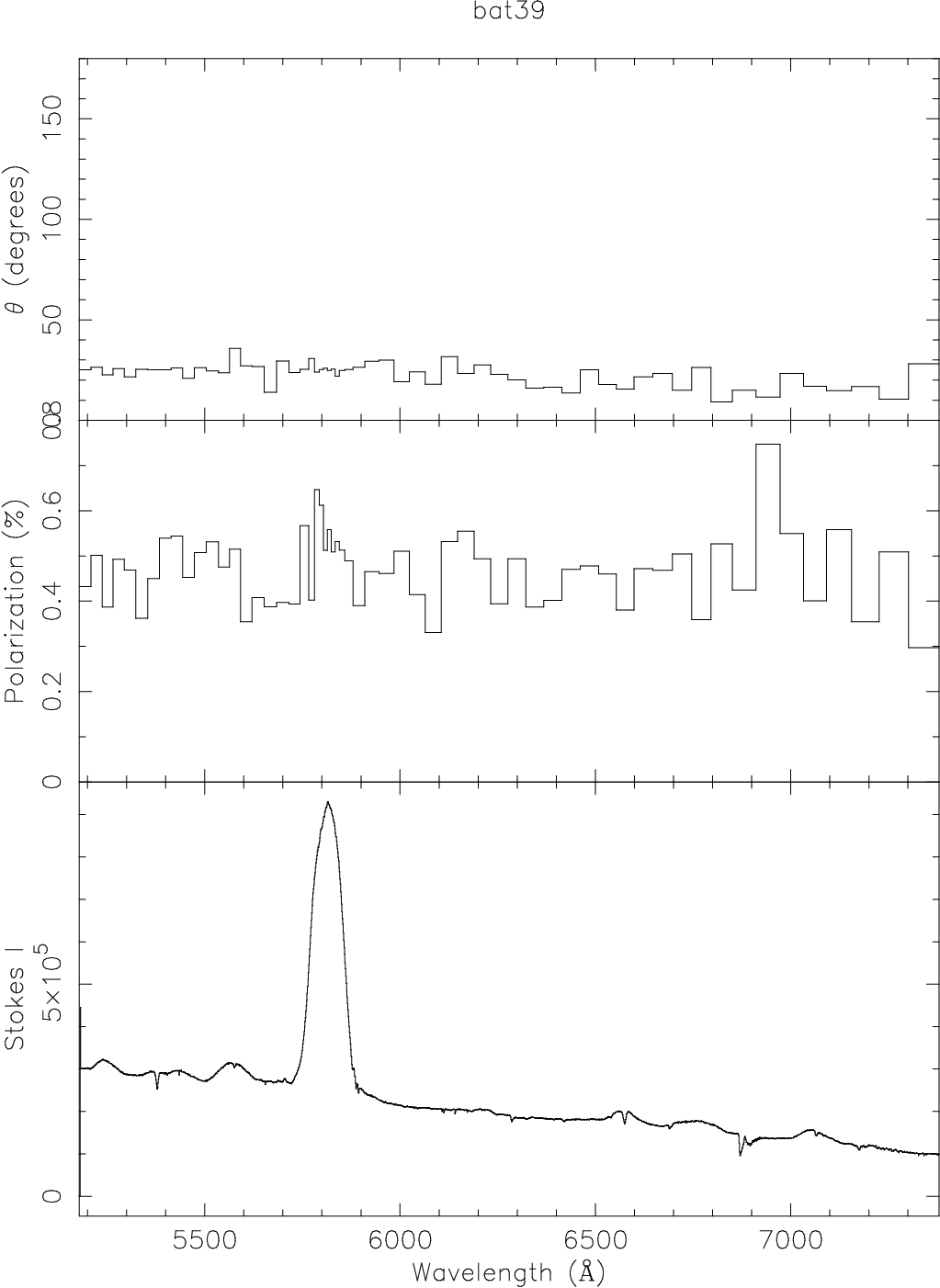}
\epsfxsize=0.33\textwidth\epsfbox{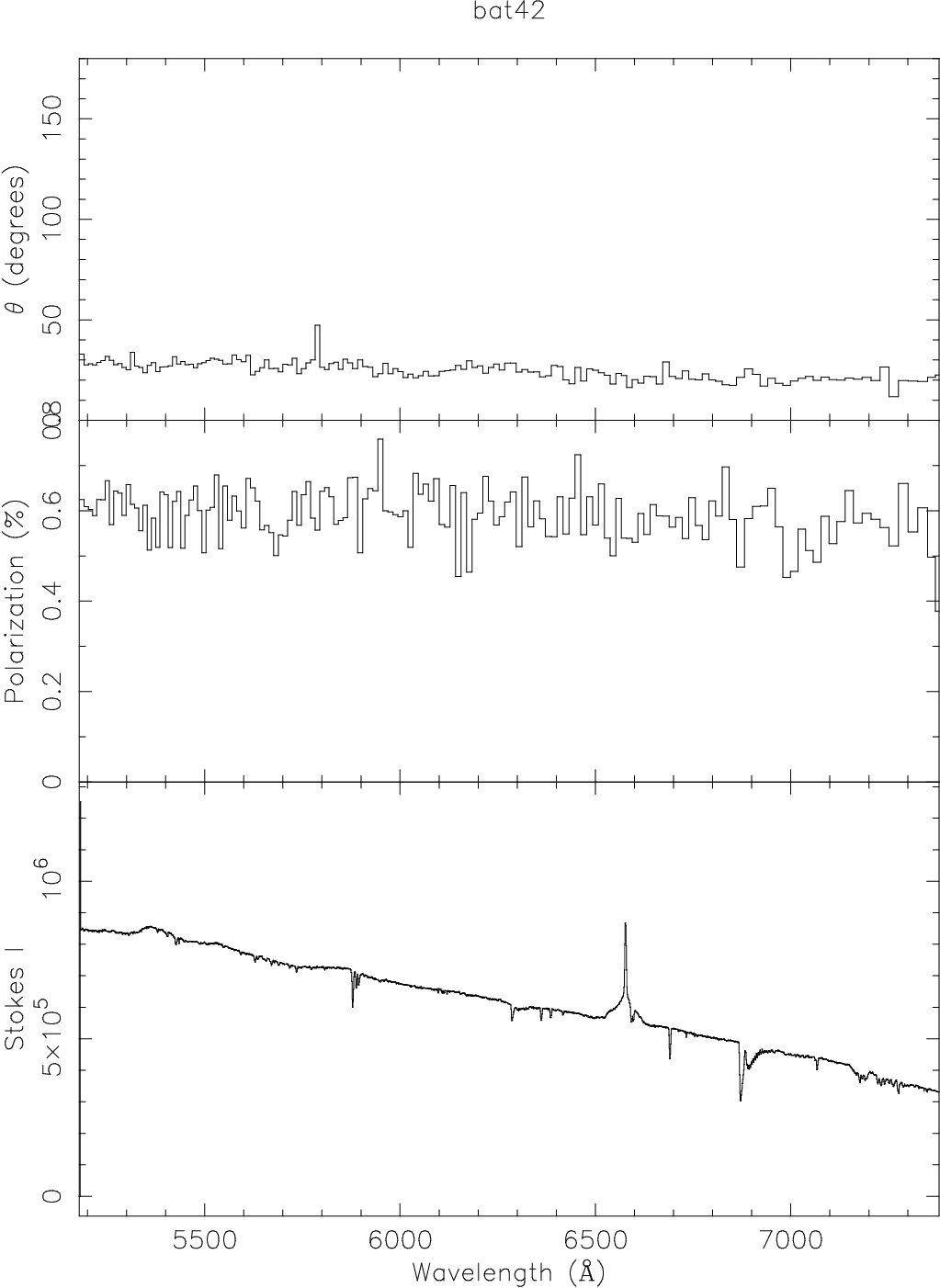}
\epsfxsize=0.33\textwidth\epsfbox{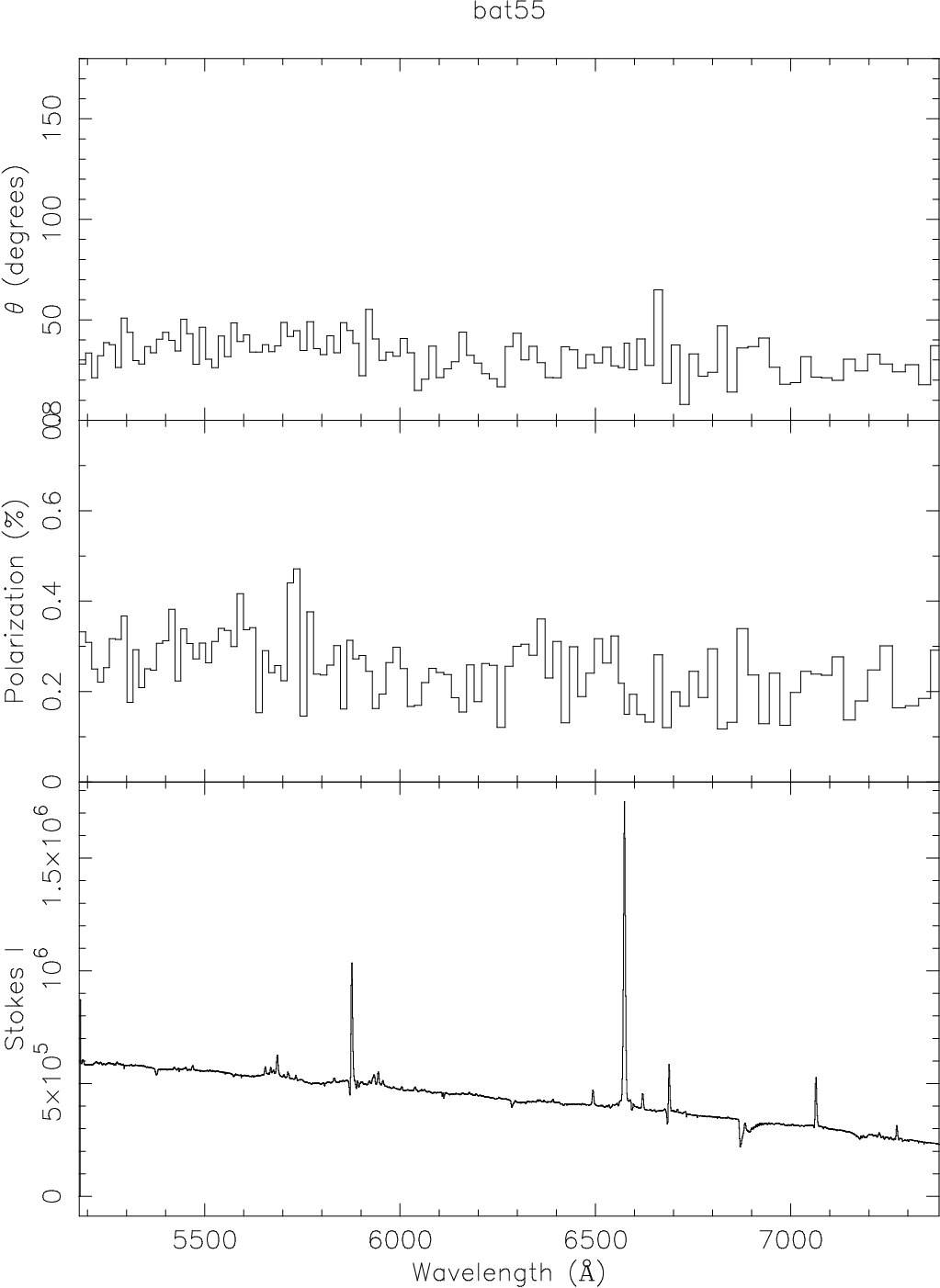}
}
\mbox{
\epsfxsize=0.33\textwidth\epsfbox{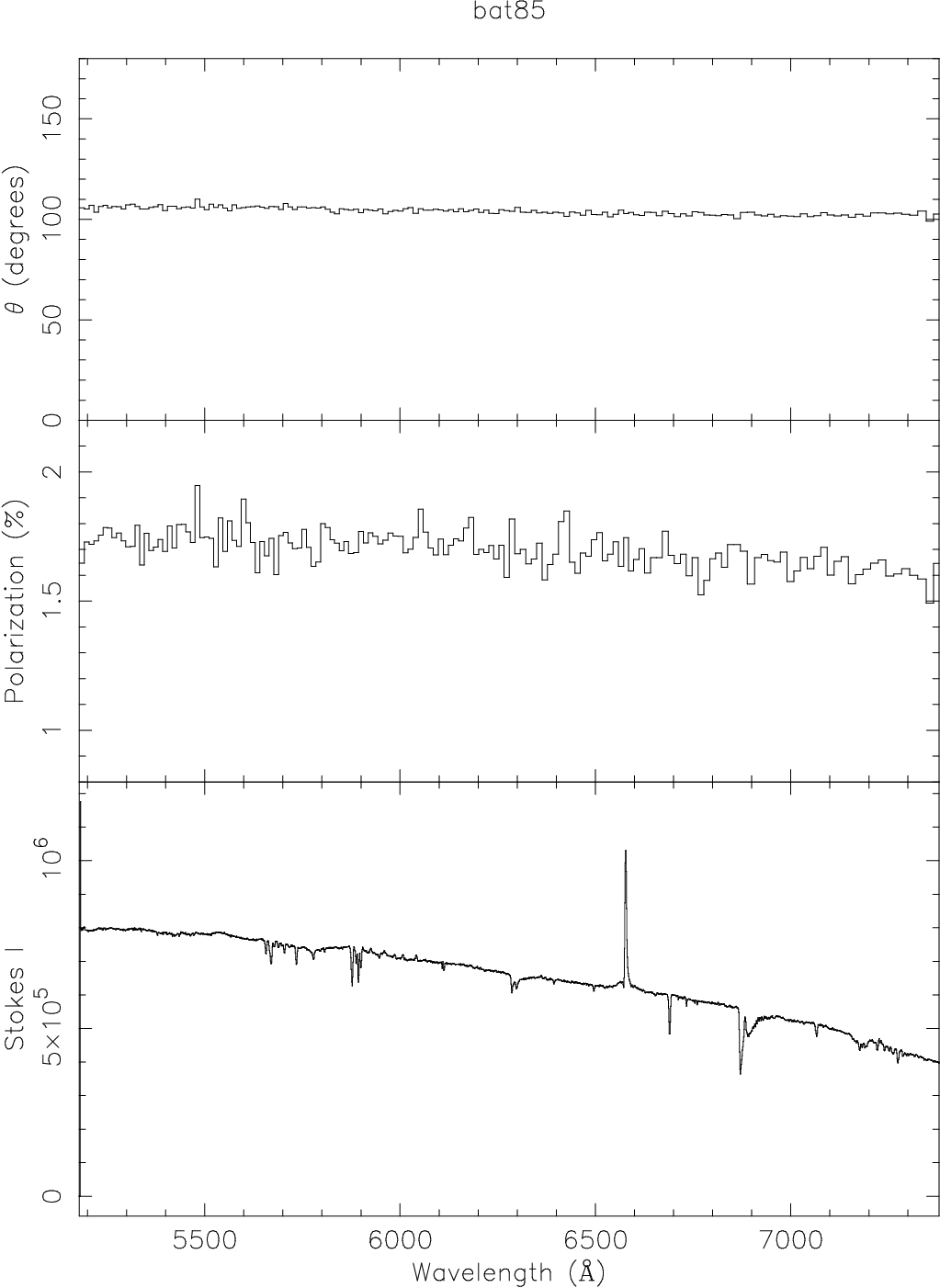}
\epsfxsize=0.33\textwidth\epsfbox{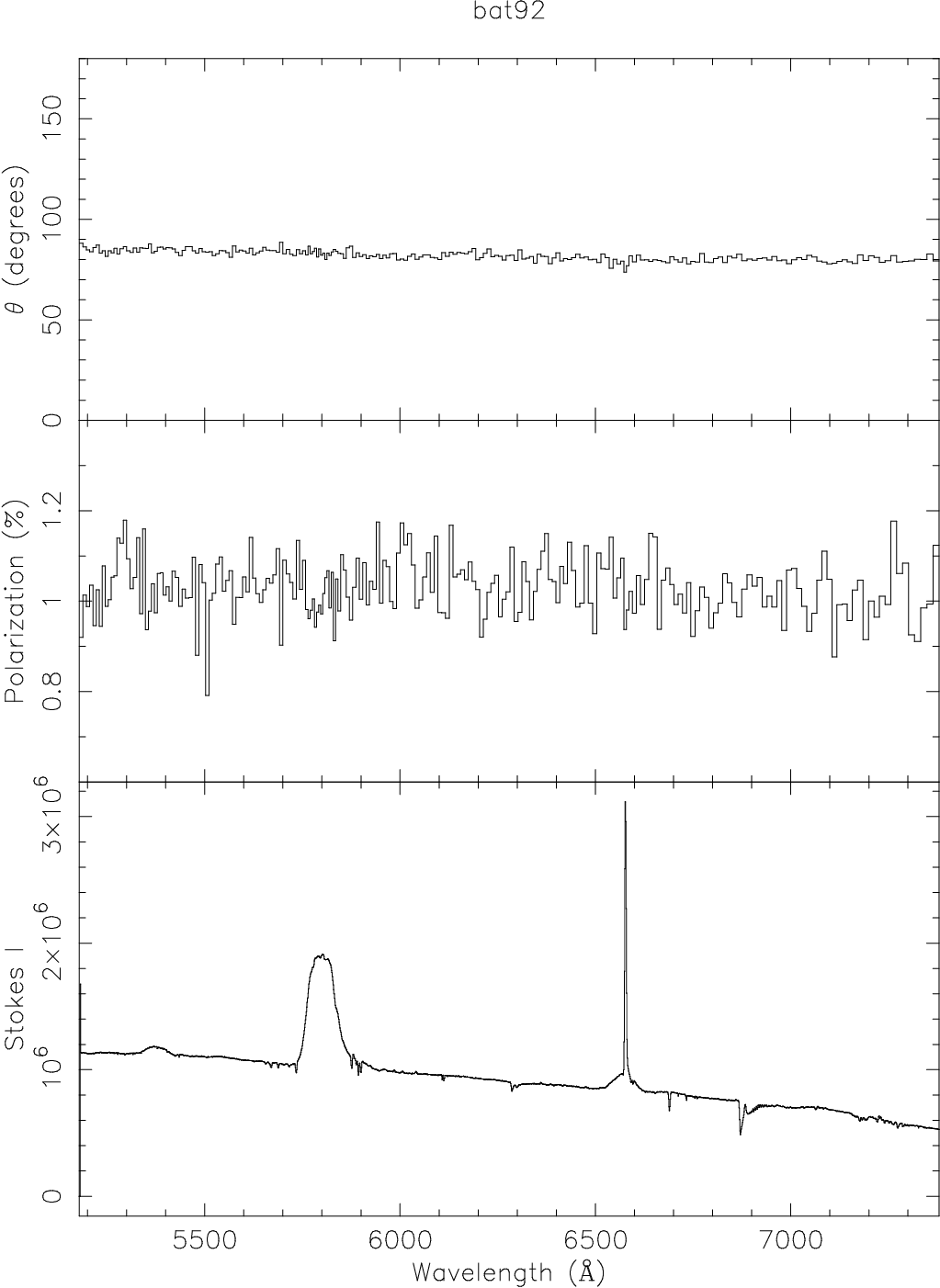}
\epsfxsize=0.33\textwidth\epsfbox{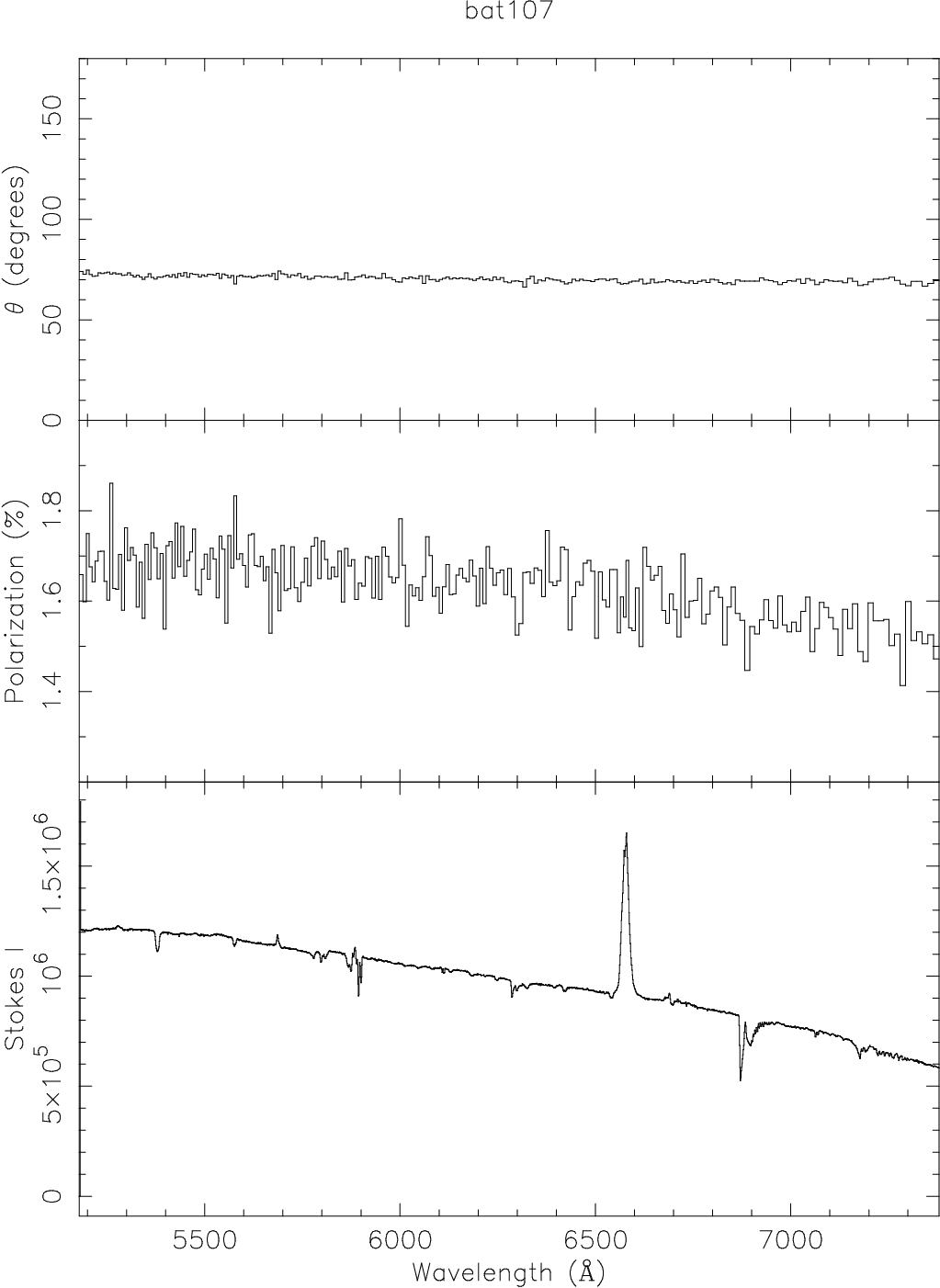}
}
\caption{Polarization triplots for  the WR stars
that do not show line depolarizations.}
\label{f_line}
\end{figure*}

\addtocounter{figure}{-1}
\begin{figure*}
\mbox{
\epsfxsize=0.33\textwidth\epsfbox{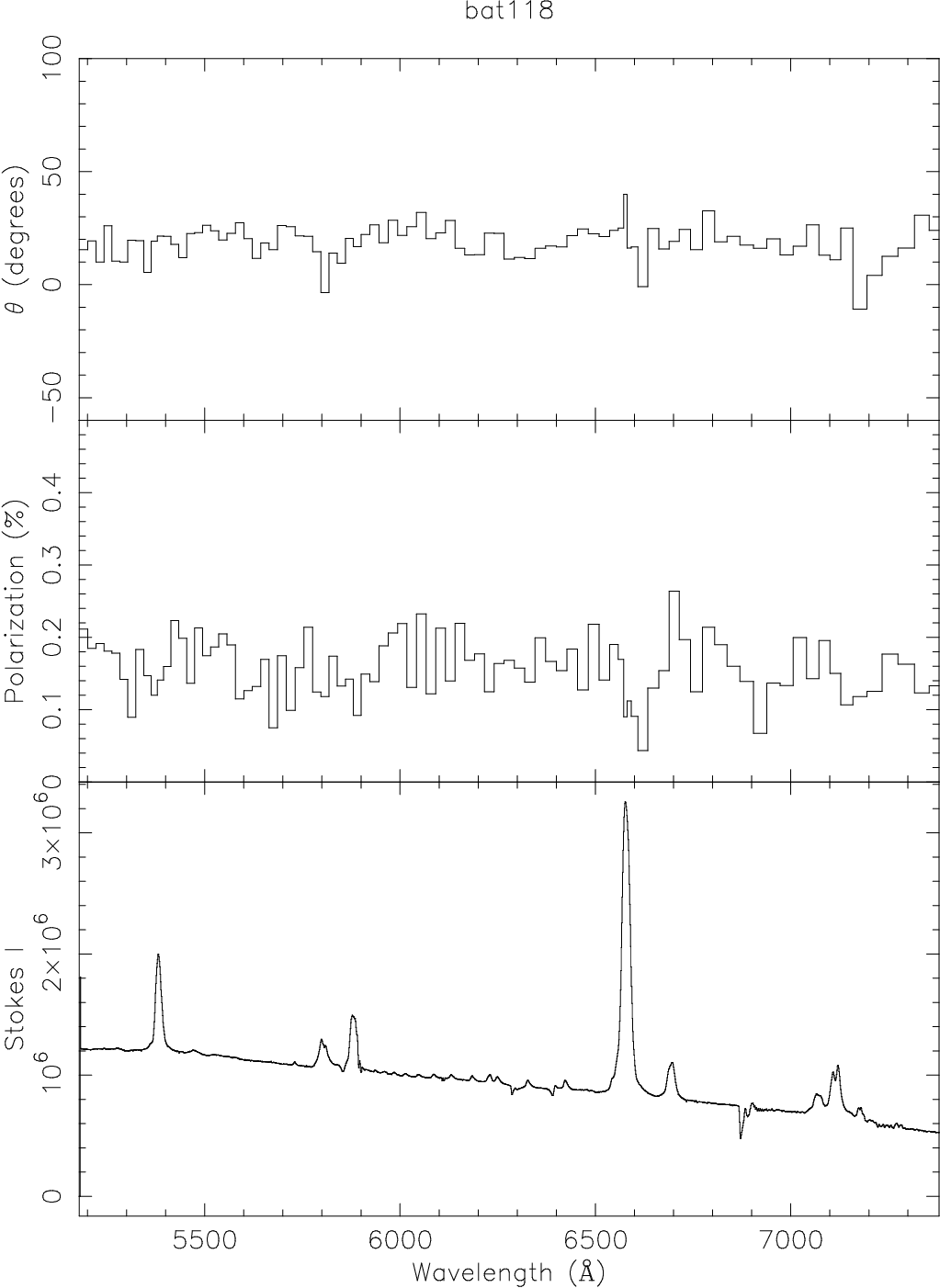}
\epsfxsize=0.33\textwidth\epsfbox{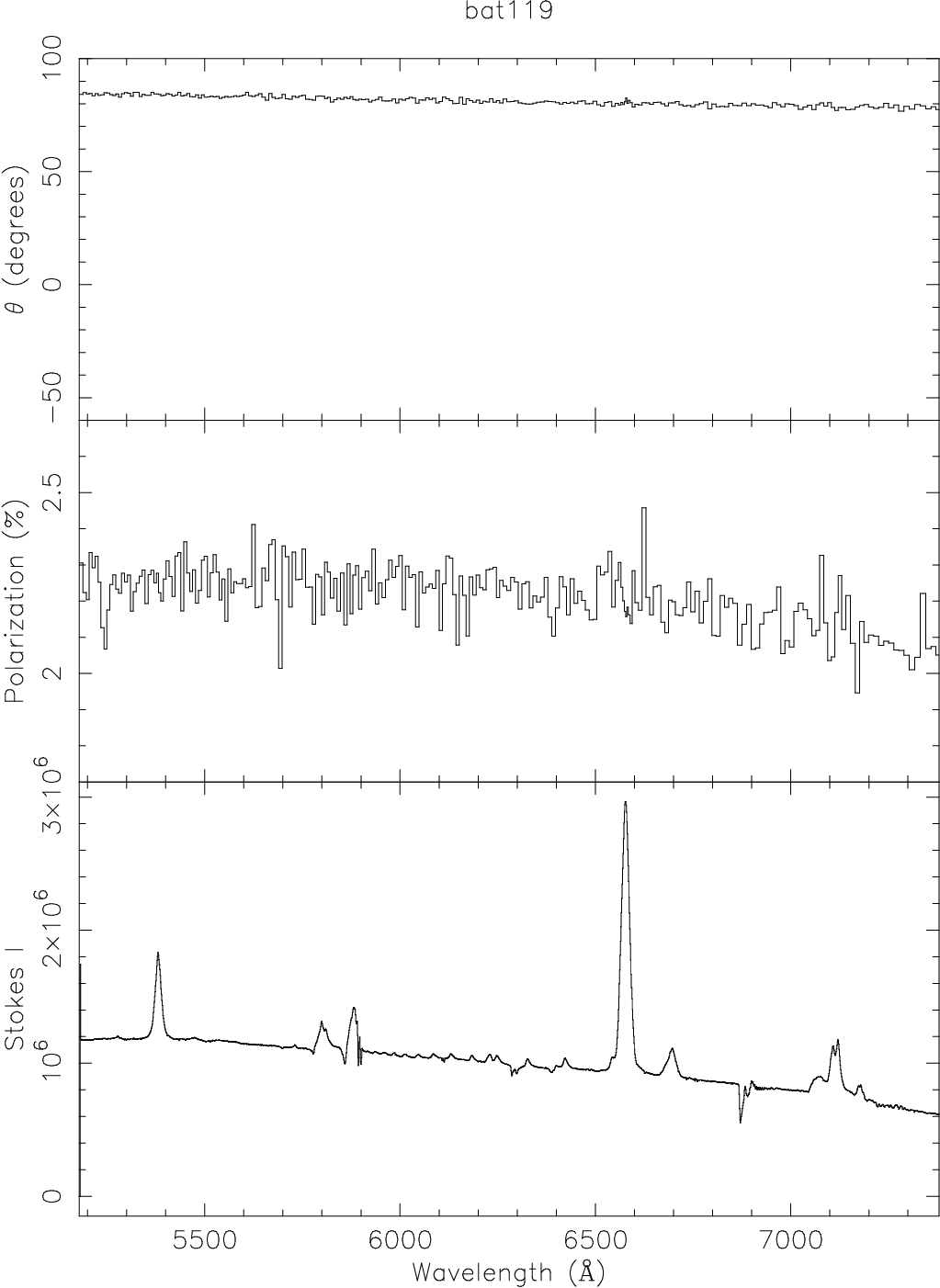}
}
\caption{Continued}
\label{f_line}
\end{figure*}

The only two objects that appear to be subject to significant intrinsic polarization (with $\ga$ 0.3\%) are 
BAT 22 and BAT 33 for which line polarizations effects can be observed across 
the He {\sc ii} line (at $\lambda$ 6560\AA) as well as at least one other line. 
For, BAT 118 we cannot definitively exclude a line effect, but because it would at most represent a marginal
detection, we count it amongst the non-detections and include it in Fig.~2.

The clearest detection is that of BAT 33. Depolarization can be observed 
in the middle panel of the left-hand side of Fig.~1 across both the He {\sc ii} lines 
at 6560\AA\ and at 5976\AA, whilst the position angle remains constant across the lines. 
The case of BAT 22, also designated R~84, is less straightforward. 
Here, the polarization increases across the  He {\sc ii} 6560\AA\ line, 
whilst the PA is also found to change. This may be due to an alignment of the interstellar and 
intrinsic polarization (see e.g. Vink et al. 2005 for the extreme case of FU~Ori). 
We can estimate the maximum amount of interstellar foreground polarization using 
the standard relation between $A_{\rm V}$ and $\%P$, $\frac{P^{\rm max}}{A_{\rm V}}$ = 3 
of Serkowski et al. (1975). Heydari-Malayeri et al. (1997) derive a relatively 
large $A_{\rm V}$ of 0.75, which could result in a maximum linear 
polarization of $\sim$2\% using the relation between ISP and reddening. 
We note that these values are maximal and highly uncertain (due 
to the uncertainties in the used Serkowski relation). In addition, we cannot 
exclude the possibility that the line itself is polarized. 
Line polarization may vary for lines formed at different positions in the wind, which 
may result in differing behaviour of the He {\sc ii} 6560\AA\ and 5976\AA\ lines. A further 
discussion on R~84, often termed a ``transitional object'' would require more 
in-depth modeling, which is beyond the scope of this paper.

Harries et al. (1998) found the Galactic line-object WR stars to be 
amongst the brightest, largest mass-loss rate systems and we now examine whether the LMC 
line-effect objects are also ``special'' in terms of their stellar parameters. 
Four (out of 13) of our LMC objects, BAT 18, BAT 55, BAT 118 and BAT 119, have been analyzed 
by Crowther \& Smith (1997) with the aim of determining the photospheric/wind parameters. 
Although BAT 22 and BAT 118 are indeed high luminosity/large mass-loss systems, so is 
BAT 119, for which we have not detected a line effect. The current sample is too small 
to assess whether wind asymmetry of LMC WR stars is restricted to large mass-loss systems.

Regarding duplicity, Foellmi et al. (2003) recently performed a detailed investigation into the multiplicity
of LMC WR stars. Unfortunately, their sample, and our brighter sample, have no objects in common. 
At this stage, it is thus not possible to assess whether the line-effect systems are binaries and 
whether the polarization effects might be due to asymmetric electron scattering around rapidly rotating single 
stars, or whether the polarization might be the result of intra-binary scattering instead. 

All in all, we put the incidence of line effects at 2 ($\pm$ 1) out of a sample 
of 13 LMC WR objects, corresponding to a line effect frequency of $\sim$ 15\%.
Equation (23) of Brown \& McLean (1977) shows that the percentage polarization scales 
as $P$ $\propto$ sin$^{2} i$, implying that we may not detect any polarization for low-inclination 
objects -- even if their winds are significantly axi-symmetric. 
Our data are of similar quality (in terms of S/N) as those in Harries et al. (1998) and
the incidence of line effects is comparably low to that of the 
$\sim$ 15-20\% of line effects detected in Galactic WR stars by Harries et al.

\section{Discussion and Conclusions}
\label{s_disc}

In terms of their polarization properties, the LMC WR stars appear to be indistinguishable 
from their Galactic counterparts, and one may wonder whether WR properties are simply the same 
at different metallicities. Recently, Marchenko et al. (2007) performed a spectroscopic 
monitoring campaign of Small Magellanic Cloud (SMC) WR stars and found their moving 
subpeaks to behave similarly in the SMC as in the Galaxy. 
Rather than due to global rotation-induced asymmetric structures, one may wonder if the WR 
polarization may also be due to wind clumping, as is believed to be the case 
for Luminous Blue Variables (Davies et al. 2005, see also Nordsieck et al. 2001). 
However, Harries et al. (1998) considered this option of inhomogeneous 
winds to account for the Galactic WR distribution of polarization levels, but found 
the statistically most significant results when a 15-20\% minority of WR stars reached measurable depolarizations 
due to flattened winds, whilst the large majority were slow rotators with spherically 
symmetric winds. The conclusion was drawn that the inferred asymmetries were only 
present for the most rapidly rotating WR stars (see also Villar-Sbaffi et al. 2006). 

Taking our results and those of Harries et al. (1998) at face value, we
may conclude that there is no significant difference
in the rotational properties of WR stars in the LMC and the Galaxy, unless the physical 
requirements to produce polarization are significantly more stringent for WR stars at LMC metallicity. 
However, 
intrinsic linear polarization {\it is} observed 
for WR stars with a range of mass-loss rates and metallicities, as well as for objects 
with much smaller mass-loss rates than WR stars (cf. classical Be stars).
It is therefore most likely that the rotational properties of LMC WR stars are 
indistinguishable from those in the Galaxy, with just a handful of objects rotating rapidly, whilst 
the majority rotate slowly.
In other words, mass loss appears to be sufficient to remove the angular momentum not only in the Galaxy, but also in 
the LMC. 

We speculate that one would need to assess WR stars at lower metallicity than the 
the $Z$ $\sim$ 0.5 $\zsun$ of the LMC, 
for the winds to be weak enough to prevent angular momentum loss.
A spectropolarimetric survey in 
a $Z$ $\sim$ 0.2 \zsun\ 
environment would be extremely useful, however the faintness of SMC WR stars 
prevents a spectropolarimetric survey of these objects -- even with an 8m-telescope. 
Future spectropolarimetric programs with {\it Extremely Large Telescopes} may 
address the question of the rotationally induced polarization properties of GRB progenitors 
more conclusively, if these instruments are equipped with the required polarization optics. 
In addition, the effect of low $Z$ wind asymmetries on 
GRB afterglow observations would be interesting (see Eldridge 2007).

Wolf \& Podsiadlowski (2007) recently suggested that GRB progenitor models that 
{\it require} a metallicity much less than $Z$ $\sim$ 0.5 $\zsun$ appear inconsistent with 
constraints from host-galaxy luminosities.
Here, the low incidence of WR polarization signals suggests that the metallicity threshold 
where significant differences in WR rotational properties 
might occur is below that of the LMC. 
Hence, we may constrain GRB progenitor models 
to an {\it upper} metallicity of that of the LMC, i.e. $Z$ $\sim$ 0.5 $\zsun$.


\begin{acknowledgements}

I would like to thank the referee, Tim Harries, for constructive comments that have 
helped improve the paper and the friendly staff at ESO/Paranal for their help.

\end{acknowledgements}

\end{document}